\documentclass{iaus}
\usepackage{graphicx}

\title[~~Outer disks of S0 galaxies] 
{Outer stellar disks \\ of lenticular galaxies}

\author[Olga K. Sil'chenko]   
{Olga K. Sil'chenko$^1$
}

\affiliation{$^1$Sternberg Astronomical Institute of the Lomonosov Moscow State University, \\ University av. 13,
119991, Moscow, Russia \\ email: {\tt olga@sai.msu.su}} 

\pubyear{2011} 
\volume{284}  
\pagerange{1--12}
\jname{The Spectral Energy Distribution of Galaxies}
\editors{R.J. Tuffs \&  C.C.Popescu, eds.}
\begin{document}

\maketitle

\begin{abstract}
By studying the stellar population properties along the radius in 15 nearby S0
galaxies, I have found that the outer stellar disks are mostly old, with the SSP-equivalent
ages of 8--15~Gyr, being often older than the bulges. This fact puts into doubt
a currently accepted paradigm that S0 galaxies have formed at $z=0.4$ by quenching star
formation in spiral galaxies.
\end{abstract}

\firstsection 
\section{Introduction}

Lenticular galaxies have been introduced by Edwin \cite[Hubble (1936)]{hubble}
as an intermediate type between ellipticals and spirals: they have large-scale stellar
disks as spirals but lack blue spiral arms and HII-regions, and they look smooth
and red as ellipticals. Now lenticulars are thought to be (trans-)formed from spirals
by removing gas and quenching star formation in their disks; also dynamical heating
is required to make the S0 disks stable against spiral wave perturbations. There are
some evidences that this transformation might take place at the redshift of $z\approx 0.4$,
within dense environments, groups or clusters, where and when suddenly the dominance of blue
(spiral?) galaxies is replaced by S0 dominance (e.g., \cite[Fasano et al. 2000]{fasano},
\cite[Wilman et al. 2009]{wilman}). But if this scenario is valid, the star formation
in the outer disks of nearby S0s should proceed only 4 Gyr ago, and we must see
intermediate-age stellar populations there. With deep long-slit spectroscopy
of nearby S0 galaxies, I study the stellar population properties 
along the radius beyond several scalelengths of their large-scale disks
to check the validity of this paradigm.

\section{Sample}

The sample consists of nearby lenticular galaxies for which
deep long-slit spectra have been obtained at the Russian 6m telescope
for the last five years in the frames of several observational programs. The main part
of the sample are edge-on lenticular galaxies selected for kinematical study
by Natalia Sotnikova which have been observed in the frame of her
observational proposal; I use here these data to derive Lick indices.
Four S0 galaxies seen moderately inclined represent a part of our sample
of nearby early-type disk galaxies -- group members whose central parts have been studied 
earlier with the Multi-Pupil Fiber Spectrograph of the 6m telescope (\cite[Sil'chenko 2006]{me06}).
The galaxies are homogeneously distributed over the luminosities,
their blue absolute magnitudes being spread from --19 to --21, and over the environment
densities. We have one galaxy (NGC~4570) in the Virgo cluster where the intracluster medium 
influence is inavoidable, and one galaxy (NGC~4111) in the Ursa Major Cluster where X-ray gas 
is not detected. Among group galaxies, NGC~524 and NGC~5353 are the central group galaxies
embedded into X-ray haloes, NGC~5308 is a member galaxy in the X-ray bright group, and NGC~502
and IC~1541 though being members of rich groups, lie outside their X-ray halo (\cite[Osmond
\& Ponman 2004]{xray}; also we have checked the archive ASCA images).
NGC~3414 is a central galaxy in the rich group undetected in X-ray. NGC~2732 is a host 
of a few faint satellites. NGC~1029, NGC~2549, and NGC~7332 are in triplets. 
By using the NED environment searcher, we have no found any galaxies
within 300 kpc off NGC~1032 and NGC~1184 so we take them as isolated field galaxies.
 
\section{Observations}

The long-slit spectral observations have been made with the focal
reducer SCORPIO (\cite[Afanasiev \& Moiseev 2005]{scorpman}) installed at the 
prime focus of the Russian 6m telescope (at the Special Astrophysical Observatory 
of the Russian Academy of Sciences). We exposed the rather narrow spectral range 
rich by absorpion lines, 4800--5500~\AA, which is quite suitable for the study
of stellar kinematics and stellar population properties. The slit width 
was one arcsecond, the spectral resolution -- about 2~\AA.  The CCD 2k$\times$2k
and 2k$\times$4k were used as detectors, and the scale along the slit was
0.36 arcsec per pixel. The slit length is about 6 arcminutes so at 
the edges of the slit we can take the sky background to subtract from the 
galaxy spectra. Inhomogeneties of optics transparency and spectral resolution
along the slit are checked with the twilight exposures. The Lick index
system was calibrated by observing the standard Lick stars (\cite[Worthey et al. 1994]{woretal}) 
with the same instrumental setup as the galaxies were observed.
We calculated the Lick indices H-beta, Mgb, Fe5270, and Fe5335 along
the slit up to several scalelengths of exponential disks.  To estimate the
radial variations of the SSP-equivalent ages, metallicities and  magnesium-to-iron
ratios of the stellar populations, I confronted the measured Lick indices to the models 
of old stellar populations by \cite[Thomas et al. (2003)]{thomod}.

\begin{figure}[b]
\begin{center}
\includegraphics[width=3.4in]{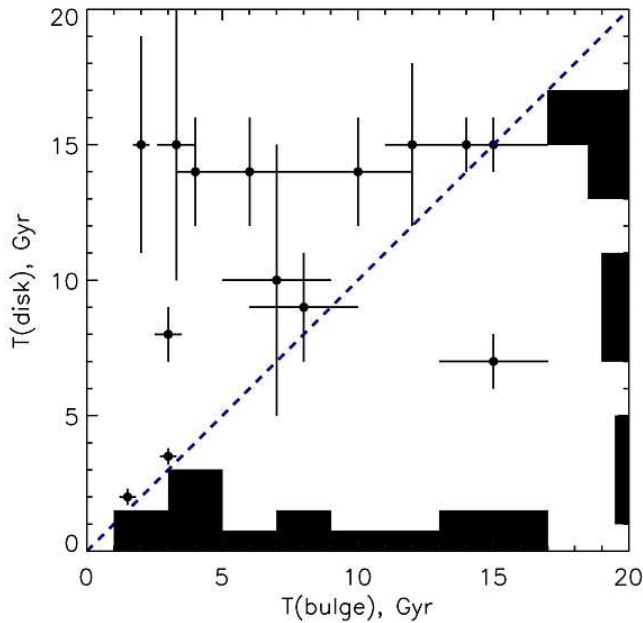} 
\caption{Comparison between the SSP-equivalent ages of the bulges, at $0.5r_e$, and of
the large-scale disks for the sample lenticulars. The dashed line is the line of equality;
the filled histograms attached to the axes characterize the distributions of the bulge
and disk ages.}
\label{fig1}
\end{center}
\end{figure}

\section{Results and Discussion}

I have measured the Lick indices in 15 S0s along the radius 
beyond 2--4 exponential scalelengths of their
disks, and have estimated the ages and abundances of
the stellar populations for the bulges and for the 
disks. The bulges have solar metallicities
or higher, and the disks have solar metallicity or lower. The disks
demonstrate very high magnesium-to-iron ratios of their
stellar populations so they cannot be descendants of spiral galaxies
with prolonged star formation. The SSP-equivalent ages of
the disks are mostly old -- 60\%\ of the sample demonstrate the
ages older than 10~Gyr, -- and almost always (except one case) 
larger than the SSP-equivalent ages of their
corresponding bulges (Fig.~\ref{fig1}). 

The parameters of the stellar populations of the disks do not show any
correlation with the disk luminosities or masses.
The only correlations found are those with the photometric disk scaleheight
taken from the decomposition results by \cite[Mosenkov et al. (2010)]{mosenkov}.
The whole scaleheight range is from 0.3 kpc (found in the disks with the ages of 2--3~Gyr 
and [Mg/Fe]$< +0.2$) to 0.6--0.9 kpc (found mostly in the disks with the ages $>10$~Gyr                                            
and [Mg/Fe]$>+0.3$). All the old disks of our sample galaxies are so 
thick disks, while the only two young outer disks, those of NGC~4111 and NGC~7332, 
are certainly thin disks. The disk of NGC~2732, with its SSP-equivalent age of 8~Gyr 
and the scaleheight of 0.5 kpc, is halfway between thin and thick disks.

Many arguments evidence for S0s and spiral galaxies being relatives, but this time the
thing looks like S0s are progenitors of spirals, opposite to what is thought before.
Indeed, if we compare stars of the thick disk of our own Galaxy with the thick outer 
stellar disks of S0s studied here we will see full resemblance: the ages $>10$ Gyr, 
[Mg/Fe]$>+0.2$, the total metallicity [Z/H] is between 0.0 and --0.7 
(such parameters of the thick disk of our Galaxy are found by e.g. \cite[Bernkopf 
\& Fuhrmann 2006]{bernfuhr} or by \cite[Schuster et al. 2006]{schuster}). 
So if one provides fresh cold gas accretion into the disks of these S0s, after several 
Gyrs of star formation we would get a typical spiral galaxy, with the thick old stellar 
disk and thin younger stellar disk. I propose now the following evolutionary sequence.
All disk galaxies were S0s immediately after their birth; later, at $z<1$, some of them acquired
cold gas accretion sources -- these became spirals, -- and some of them failed to find
such sources -- those remained lenticulars. In large cluster-size and group-size dark
haloes, there are little chance to find external sources of cold gas accretion, due to hot 
gas effect, -- so in nearby clusters the dominant disk-galaxy population is S0s. 
The open question remains what can be these sources of cold-gas prolonged accretion  
-- they may be cosmologically motivated filaments (\cite[Dekel \& Birnboim 2006]{dekel}) 
or rich systems of irregular-type dwarf satellites which have been merging with the host 
one after another.

\end{document}